\documentclass[10pt,letterpaper]{article}
\usepackage[latin9]{inputenc}
\usepackage{amsmath}
\usepackage{amssymb}
\usepackage{graphicx}
\usepackage{esint}
\usepackage{amsfonts}
\usepackage{opex3}
\usepackage{cite}
\usepackage{pdfpages}

\begin{document}

\title{Efficient coupling between dielectric waveguide modes and exterior
plasmon whispering gallery modes}
\author{Chen-Guang Xu,$^{1}$ Xiao Xiong,$^{1}$ Chang-Ling Zou,$^{1,2}$
Xi-Feng Ren,$^{1,*}$ and Guang-Can Guo$^{1}$}
\address{
$^1$ Key Lab of Quantum Information, University of Science and Technology of
China, Hefei 230026, P. R. China\\
$^2$ clzou321@ustc.edu.cn}
\email{$^{*}$ renxf@ustc.edu.cn}

\begin{abstract}
Inefficient coupling between dielectric guided mode and plasmon mode has
been overlooked in the past. The mechanism in it is essentially different
from the conventional coupling between dielectric modes. Based on
qualitative theoretical analysis, we proposed two methods to strengthen the
coupling between dielectric waveguide modes and exterior plasmon whispering
gallery modes. One is using a U-shaped bent waveguide to break the adiabatic
mode conversion process, and the other is to render the dielectric mode of
higher-order to reach phase matching with plasmon mode. Both the
transmission spectrum of waveguide and the energy spectrum of cavity
demonstrate that the coupling efficiencies can be greatly improved. These
simple configurations are potential for wide applications, for example,
tunable integrated optical devices, nanolasers and sensors.
\end{abstract}




\section{Introduction}

Photonic devices based on surface plasmon polaritons (SPPs) have been
extensively investigated recently \cite{1}. SPP as collective oscillation of
free electrons in the metal surface, has many excellent properties, such as
sub-wavelength confined local field \cite{2,3} and strong polarization
selectivity \cite{16}. It perfectly combines electrons and photons together
in the nanoscale. The strong concentration of electromagnetic field greatly
benefits light-matter interactions, driving the development of highly
sensitive sensors\cite{4,4-1}, nanolasers \cite{4-2,4-3} and photovoltaic
cells \cite{4-4}.

Plasmonic nanostructures such as cavities \cite{cav} and waveguides \cite%
{xiongrev}, which can confine light in very small volume and greatly enhance
the light-matter interaction, have attracted increasing attention. For
example, the exterior whispering gallery mode (WGM) in plasmonic resonator
\cite{Exterior1,Exterior2,7}, which has the excellent properties like high
quality (Q) factor and small mode volume, permits strong light-matter
interactions since most of the light is confined outside the cavity. Such
exterior WGMs hold great potential for various applications, such as highly
sensitive bio-sensors. However, unlike dielectric components, the
applications of plasmonic components are greatly limited due to the
inefficient excitation and collection of SPPs \cite{4-5}, though optical
taper has been used in experimentally coupling. This is because the
excitation and collection of SPPs with dielectric waveguide are essentially
different from traditional dielectric WGMs, which have been overlooked.

In this Letter, we theoretically studied the coupling between exterior WGMs
in metallic microresonator and guided modes in dielectric waveguide. It is
shown that the underlying mechanism of this type of coupling is
significantly different from that of coupled dielectric waveguides, where
the traditional perturbation coupled mode theory is applied. Based on our
theoretical analysis, we proposed two methods that can greatly raise the
coupling efficiency between the dielectric waveguide and metallic resonator:
One is to break the adiabatic conversion condition during mode propagation,
realized by altering the geometry of the waveguide; the other one is
utilizing guided mode of higher-order in dielectric waveguide to reach phase
matching. Furthermore, we calculated energy spectrum and transmission
spectrum for these two schemes , which demonstrate the feasibility and show
good performance. Since the mechanism of the coupling between dielectric
mode and SPPs is general, these two methods can also be applied to the
coupling between dielectric and plasmonic waveguides, in which case the
perturbation theory is invalid.

\section{Model}

In the following studies, we investigated the coupling by numerically
solving the Maxwell equations with finite element method (COMSOL
Multiphysics 4.3). We focus on the working wavelength $\lambda =532.8\,%
\mathrm{nm}$ (frequency $f=563\,\mathrm{T}\mathrm{Hz}$), with the relative
permittivity of dielectric (silica) being $\varepsilon _{silica}=2.134$.

\begin{figure}[tbp]
\centerline{\includegraphics[width=0.9\columnwidth,natwidth=850,natheight=600]{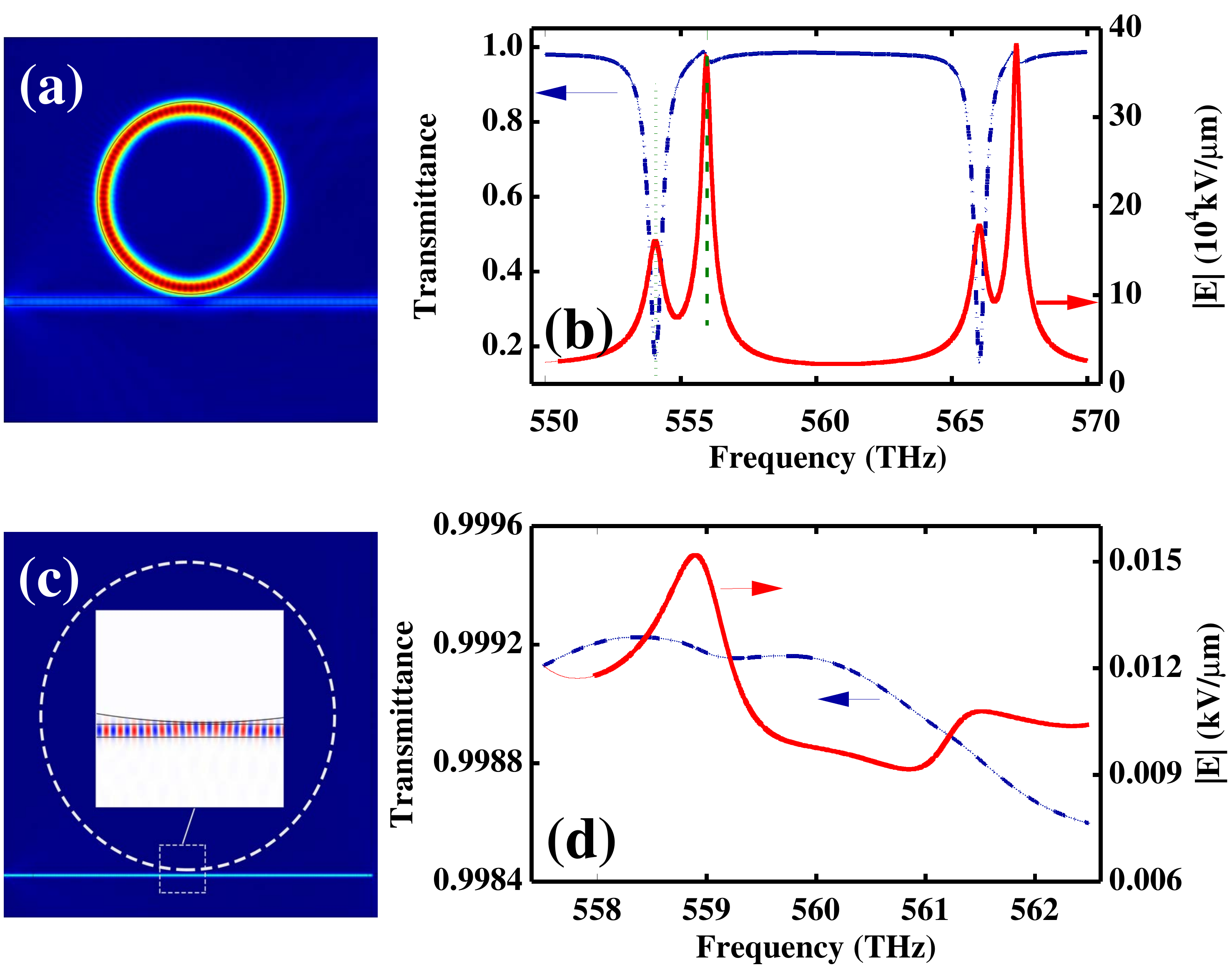}}
\caption{(a) Coupling between dielectric waveguide and dielectric cavity,
with waveguide width $0.26\,\mathrm{\mathrm{\protect\mu }m}$, cavity radius $%
3\,\mathrm{\mathrm{\protect\mu }m}$, minimum gap $100\,\mathrm{nm}$. (b)
Transmission spectrum of waveguide and the spectrum of normalized energy
inside the cavity, corresponding to the configuration in (a). (c) Coupling
between dielectric waveguide and plasmonic cavity, with waveguide width $%
0.36\,\mathrm{\mathrm{\protect\mu }m}$, cavity radius $20\,\mathrm{\mathrm{%
\protect\mu }m}$, minimum gap $50\,\mathrm{nm}$. White dashed line indicates
the outline of silver cavity. Inset: enlarged picture of the coupling
region. (d) Transmission spectrum of waveguide and the spectrum of
normalized energy , corresponding to the configuration in (c).}
\end{figure}

As an example, we started with the conventional coupling between dielectric
waveguide and dielectric cavity \cite{8}, as shown in Fig. 1(a).
Electromagnetic wave with magnetic field perpendicular to the plane
constituted by waveguide and cavity (TM mode) is loaded at the left facet of
the waveguide. And the input power is $1\,\mathrm{\mathrm{\mu }W/m}$, which
is actually the linear density of input power, since our simulations are
based on two-dimensional model. For single cavity mode, the steady state
cavity field $E_{s}$ and transmitted field $E_{t}$ can be obtained from the
input-output relation \cite{Qua-Opt} as
\begin{equation}
E_{s}=-\frac{\sqrt{2\kappa _{1}}}{i(\omega -\omega _{c})-\kappa _{0}-\kappa
_{1}}E_{i}\,,
\end{equation}%
\begin{equation}
E_{t}=-\frac{i(\omega -\omega _{c})-\kappa _{0}+\kappa _{1}}{i(\omega
-\omega _{c})-\kappa _{0}-\kappa _{1}}E_{i}\,,
\end{equation}%
where $E_{i}$ and $\omega $ are the field and frequency of input light, $%
\omega _{c}$, $\kappa _{0}$ and $\kappa _{1}$ are frequency, intrinsic loss
(including radiation, absorption and scattering losses) and external
coupling loss (coupling to the waveguide) of the cavity mode. Figure 1(b)
displays the normalized transmission spectrum ($T=\left\vert
E_{t}/E_{i}\right\vert ^{2}$) of waveguide and the spectrum of normalized
cavity field $\left\vert E\right\vert =\left\vert E_{s}/E_{i}\right\vert $.
The transmission spectrum contains symmetric Lorentzian dips and Fano-type
line-shapes, which correspond to the peaks appeared in energy spectrum and
arise from the coupling of guided mode to the WGMs in microcavity. As shown
by the field distribution in Fig. 1(a), energy is effectively coupled into
the cavity when on resonance ($\omega =\omega _{c}$). The deep dips in
transmission spectrum (dotted line in Fig. 1(b)) correspond to the 2nd-order
radial mode working near the critical coupling condition ($\kappa
_{1}\approx \kappa _{0}$), while the Fano-like transmission (dashed line in
Fig. 1(b)) is brought about by the fundamental mode working in the
over-coupling regime ($\kappa _{1}\gg \kappa _{0}$).

Afterward, the coupling between dielectric waveguide and metal (silver)
resonator is studied, as shown in Fig. 1(c). Here, the input power is about $%
1\,\mathrm{\mathrm{\mu}W/m}$, and the relative permittivity of silver is $%
\varepsilon_{silver}=-9.16+0.016i$. The very small imaginary part of $%
\varepsilon_{silver}$, which is possible at low temperature \cite{T}, is set
to reduce the Ohmic loss of silver and to raise the Q factor of the metal
cavity. In contrast to the dielectric cavity case, the transmission spectrum
and the energy spectrum for silver cavity coupling do not show obvious
resonance, as displayed in Fig. 1(d). The exterior (plasmonic) WGMs can not
be excited by the dielectric waveguide efficiently (Fig. 1(c)), since
the coupling strength is rather small compared to its intrinsic loss ($%
\kappa_{1}\ll\kappa_{0}$).

\section{Analysis}

In order to improve the coupling efficiency between dielectric guided mode
and exterior WGMs, we should firstly understand the underlying mechanism of
the coupling process. Actually, for the dielectric cavity case, light is
well confined in dielectric waveguide and cavity, with only a minor portion
of evanescent field overlapped with each other. Based on the perturbation
theory, the coupling strength is proportional to the integral of the overlap
between their wave functions \cite{9,10}. However, for metal cavity case,
energy of exterior WGMs is mostly distributed outside the cavity. When the
waveguide approach the cavity, the field overlap is very large and the field
distribution is greatly altered. Thus, the coupling between dielectric
guided modes and plasmon mode can no longer be analyzed with perturbation
theory.

\begin{figure}[tbp]
\centerline{ \includegraphics[width=1.2\columnwidth,natwidth=850,natheight=400]{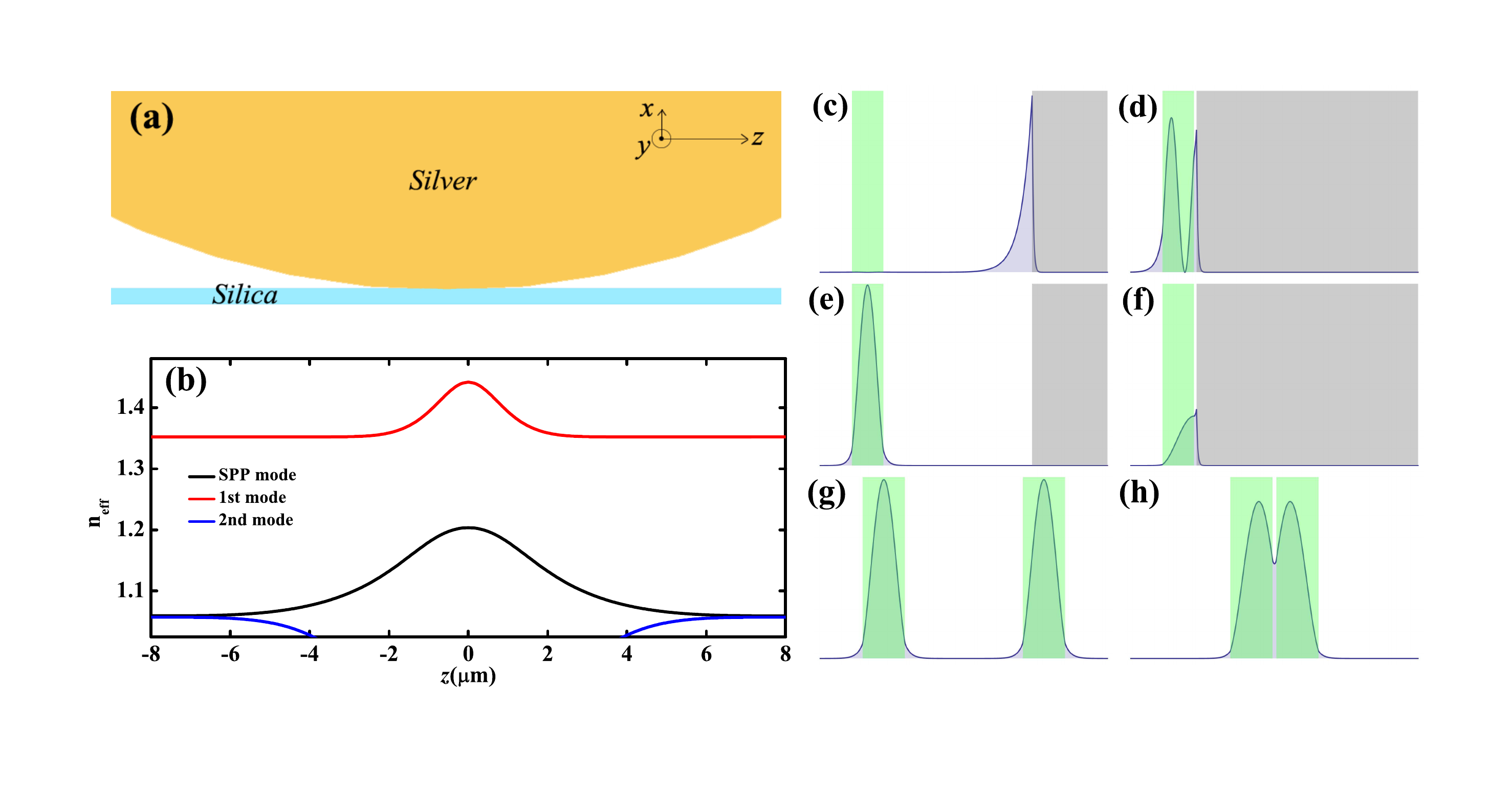}}
\caption{(a) Schematic illustration of the coupling between dielectric
waveguide and metal microresonator, with waveguide tangent to the cavity.
(b) $n_{eff}$ of eigenmodes, which is dependent on the gap between waveguide
and cavity, at each point $z$. (c)-(d) For metal cavity coupling, the
normalized electric field profiles of the eigen plasmon mode at $z=-8$ $%
\mathrm{\mathrm{\protect\mu}m}$ (c) and $z=0$ (d), respectively. The green
and gray regions represent dielectric waveguide and metal cavity,
respectively. (e)-(f) For metal cavity coupling, the normalized electric
field profiles of the eigen dielectric mode at $z=-8$ $\mathrm{\mathrm{%
\protect\mu}m}$ (e) and $z=0$ (f), respectively. (g)-(h) For dielectric
cavity coupling, the electric field profiles of the eigen dielectric mode at
$z=-8$ $\mathrm{\mathrm{\protect\mu}m}$ (g) and $z=0$ (h), respectively.}
\end{figure}

To further analyze the difference therein, we started from solving the
hybrid plasmon-dielectric modes with the dielectric waveguide coupled to
silver microresonator, the schematic of which is illustrated in Fig. 2(a).
Note that, $\kappa_{1}$, which determines the cavity mode, is dependent on
the energy transfer when waveguide and cavity draw close to each other.
Thus, we focus on the coupling region without worrying about the details of
cavity. We calculated the effective refractive indices $n_{eff}$ of
eigenmodes supported by dielectric waveguide and metal cavity at the
cross-section of point $z$, which are related to the gap between waveguide
and cavity, as shown in Fig. 2(b). The red and black curves correspond to
guided dielectric mode and plasmon mode, respectively. As shown by the
electric field profiles of eigenmodes in Figs. 2(c)-(f), both dielectric
mode and plasmon mode have been greatly changed due to mode hybridization.
When the gap between waveguide and cavity is large, dielectric mode and
plasmon mode are independent, and are confined in the waveguide and around
the cavity, respectively (Figs. 2(c) and 2(e)). But when the gap minifies,
the overlap between dielectric and plasmon modes is so large that the two
modes are hybridized and form completely distinct modes (Figs. 2(d) and
2(f)). This point is also reflected in Fig. 2(b): the three eigenmodes
evolve to be only two, as the light propagates towards the origin. Hence,
this situation differs a lot from the dielectric cavity case, where new
normal modes are linear superpositions of the two dielectric modes according
to perturbation theory (Figs. 2(g) and 2(h)).

Then, with the $n_{eff}$ of eigenmodes, mode propagation during the coupling
region can be obtained. Similar to the time evolution of quantum states, the
evolution of modes in the waveguide follows Schrödinger equation \cite%
{STIRAP2,STIRAP3}, and can be expressed as
\begin{equation}
i\frac{\partial}{\partial z}\left\vert \varphi(z)\right\rangle
=H(z)\left\vert \varphi(z)\right\rangle ,
\end{equation}
where $H(z)$ is the Hamiltonian as a function of $z$ due to the variation of
the gap between waveguide and cavity along $z$ axis, and $\left\vert
\varphi(z)\right\rangle $ is the instantaneous eigenfunction at the point $z$%
. According to Ref. \cite{11}, we obtained the coupling dynamic equations
\begin{equation}
\frac{\partial}{\partial z}a_{k}(z)=-\left\langle \varphi_{k}(z)\right\vert
\frac{\partial}{\partial z}\left\vert \varphi_{k}(z)\right\rangle
a_{k}(z)-\sum_{m\neq
k}g_{km}(z)e^{i\int(\beta_{m}(z)-\beta_{k}(z))dz}a_{m}(z).  \label{eq:2}
\end{equation}
Here, $\left\vert \varphi_{k(m)}(z)\right\rangle $ is the eigenmode of $H(z)$%
, while $a_{k(m)}(z)=\left\langle
\varphi_{k(m)}(z)\mid\varphi(z)\right\rangle $ is the corresponding
coefficient, $g_{km}(z)=\left\langle \varphi_{k}(z)\right\vert \frac{\partial%
}{\partial z}\left\vert \varphi_{m}(z)\right\rangle $ is the coupling
strength between eigenmodes $k$ and $m$, and $\beta_{m(n)}(z)=n_{eff}(z)2%
\pi/\lambda$ is the propagation constant of the eigenmode. The first term in
Eq. \ref{eq:2} is the Berry phase, which does not play a significant role in
the coupling process since its integral equals to zero. The second term in
Eq. \ref{eq:2} stands for the coupling between eigenmodes ($\left\vert
\varphi_{k(m)}(z)\right\rangle $, rather than the eigenmodes of separate
waveguides above), which reveals the requirements for efficient mode
conversion. The first factor is the coupling strength $g_{km}(z)$, another
one is the exponent $e^{i\int(\beta_{m}(z)-\beta_{k}(z))dz}$ which stands
for phase.

Since $H(z)$ is dependent on the gap between waveguide and cavity $d$ (with
the minimum gap denoted as $d_{0}$), $g_{km}(z)$ can be rewritten as $%
g_{km}(z)=g_{km}(d)\frac{\partial d}{\partial z}$, where $g_{km}(d)=\frac{%
\left\langle \varphi_{k}(d)\right\vert \frac{\partial H(d)}{\partial d}%
\left\vert \varphi_{m}(d)\right\rangle }{\beta_{k}(d)-\beta_{m}(d)}$. As
waveguide approaches the cavity, $\frac{\partial d}{\partial z}\approx-\sqrt{%
\frac{d}{2R}}$, with $R$ being the cavity radius. The slowly varying gap ( $%
R\gg\lambda$) and the large mismatching in propagation constant ($%
\beta_{k}(d)-\beta_{m}(d)$) (Fig. 2(b)) give rise to the adiabatic
conversion \cite{14}. That is to say, energy in dielectric waveguide is
coupled into metal resonator as plasmon mode, and then is converted back to
waveguide as dielectric mode. And the field distribution in Fig. 1(c) has
shown that little energy can be coupled into silver cavity with a straight
waveguide. Therefore, we need to break the adiabatic mode evolution (1) by
engineering $d$ to vary faster or (2) by phase matching. It's worth noting
that the breaking adiabatic evolution here is a fast varied process and is
essentially different from Landau-Zener transition between different states
at the anti-crossing of energy level \cite{LZ1,LZ2}, which still applies to
perturbation theory.

Inspired by the analysis above, we proposed two methods that could enhance
the coupling from silica waveguide to silver cavity. Firstly, we considered
the coupling between silver cavity and a U-shaped bent waveguide \cite{12,13}%
, the outer radius of which in the semicircle coupling region is $r$. Next,
with regard to the exponent factor, the effective refractive index of
dielectric mode needs to be reduced to match with plasmon mode. So we
investigated the behavior of the second-order dielectric mode in waveguide,
as shown by the blue curve in Fig. 2(b), which is pretty close to $n_{eff}$
of plasmon mode in silver cavity. Hence, we can utilize the second-order
optical mode in silica waveguide to reach phase matching with the plasmon
mode in silver cavity. In the following, we'll verify our protocols by
numerical simulations.

\section{Results and discussion}

\subsection{Bent waveguide coupler}

\begin{figure}[tbp]
\centerline{ \includegraphics[width=1\columnwidth,natwidth=850,natheight=350]{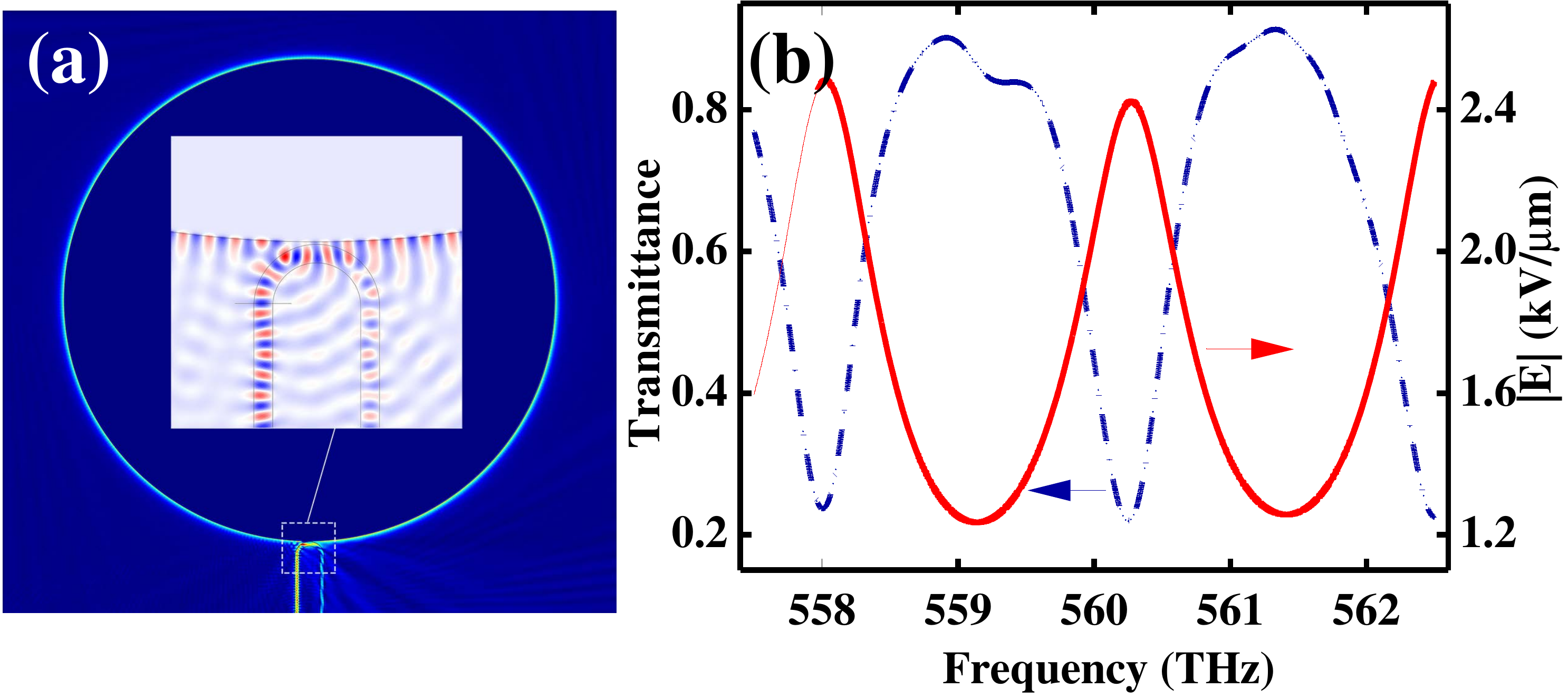}}
\caption{(a) Coupling between a U-shaped bent silica waveguide and silver
cavity, with $R=20\,\mathrm{\protect\mu m}$, $r=1.2\,\mathrm{\protect\mu m}$%
, and waveguide width being $0.36\,\mathrm{\protect\mu m}$. Inset: enlarged
picture of the coupling region. (b) Transmission spectrum of waveguide and
the spectrum of normalized energy $\left\vert E\right\vert $, with $%
d_{0}=50\,\mathrm{nm}$.}
\end{figure}

We firstly studied the coupling between a U-shaped bent silica waveguide and
silver cavity, with the input power around $1\,\mathrm{\mathrm{\mu}W/m}$. As
shown in Fig. 3(a), when the input light is on resonance with the cavity
mode, most energy is confined around the metal cavity and little is
transmitted from dielectric waveguide. This demonstrates the efficient
coupling from dielectric waveguide to metal cavity. And the effective
coupling originates from the rapidly changed gap $d$, which raises the
coupling strength between them and breaks the original adiabatic evolution.
Besides, benefiting from the properties of SPPs, energy in cavity is
extremely confined in the metal surface with an ultra-small mode volume. The
visual impression that coupling has been improved in Fig. 3(a) is further
supported by the transmission spectrum of waveguide and the energy spectrum
of cavity in Fig. 3(b). Since there is only fundamental exterior plasmon
WGMs, the spectrum shows very regular dips and peaks with equal interval.
The Q factor of cavity can be calculated from the energy spectrum as $%
Q=\omega_{c}/\Delta\omega$, with $\Delta\omega$ being the full width at half
maximum (FWHM). For the resonance at $\omega_{c}=562.6\,\mathrm{THz}$, we
obtained $Q=740$.

\begin{figure}[tbp]
\centerline{ \includegraphics[width=0.8\columnwidth,natwidth=900,natheight=500]{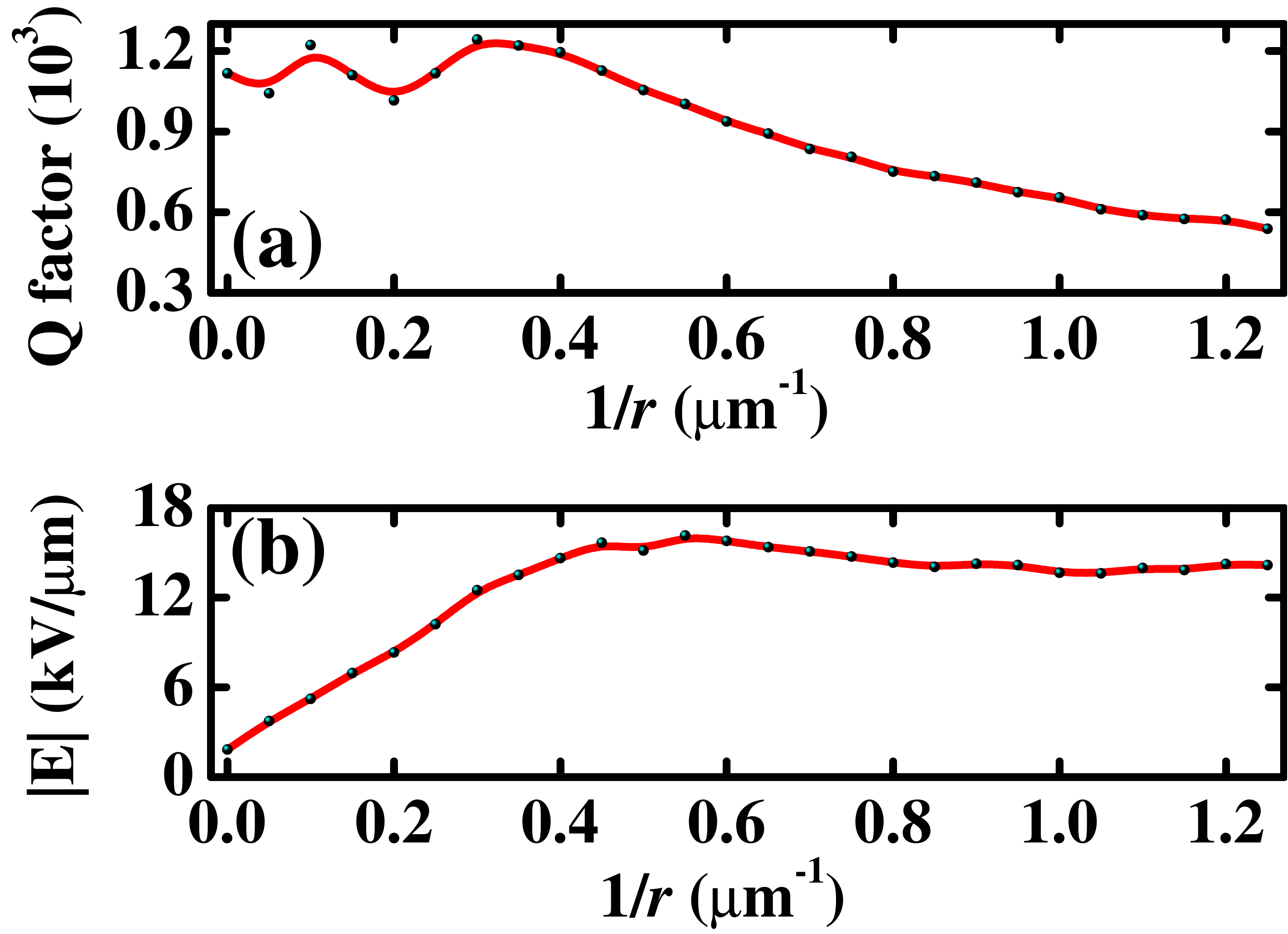}}
\caption{(a) Dependence of Q factor on $1/r$. (b) Dependence of normalized
energy $\left|E\right|$ on $1/r$.}
\end{figure}

For a better understanding, we calculated the influence of outer radius of
waveguide $r$ on the dielectric-plasmon coupling. In Fig. 4(a), the
dependence of Q factor on $1/r$ is displayed. The zero point where $1/r=0$
corresponds to a straight waveguide coupling with silver cavity, and the low
Q factor $\left(\sim1200\right)$ which is much lower than that of bare
cavity $\left(\sim3100\right)$ is due to the waveguide induced scattering
loss. The increase of $1/r$ leads to that: (1) the adiabatic change of the
gap between waveguide and metal cavity is destroyed, (2) the region that
waveguide interacts with cavity is reduced. The former effect gives rise to
the increase of coupling $\kappa_{1}$, as indicated in Fig. 4(b) that cavity
energy increases with increasing $1/r$. The later effect induces complex
behavior of cavity loss $\kappa_{0}$: as the interaction region is
shortened, the waveguide induced scattering loss decreases at first, but
very strong scattering loss is induced when the interaction region is
comparable to wavelength. At the optimal bent coupler radius $r\approx2\,\mathrm{%
\mathrm{\mu }m}$%
, $\left\vert E\right\vert $ is increased by about one order (Fig. 4(b)).

\subsection{Higher-order mode coupler}

\begin{figure}[tbp]
\centerline{ \includegraphics[width=1\columnwidth,natwidth=900,natheight=400]{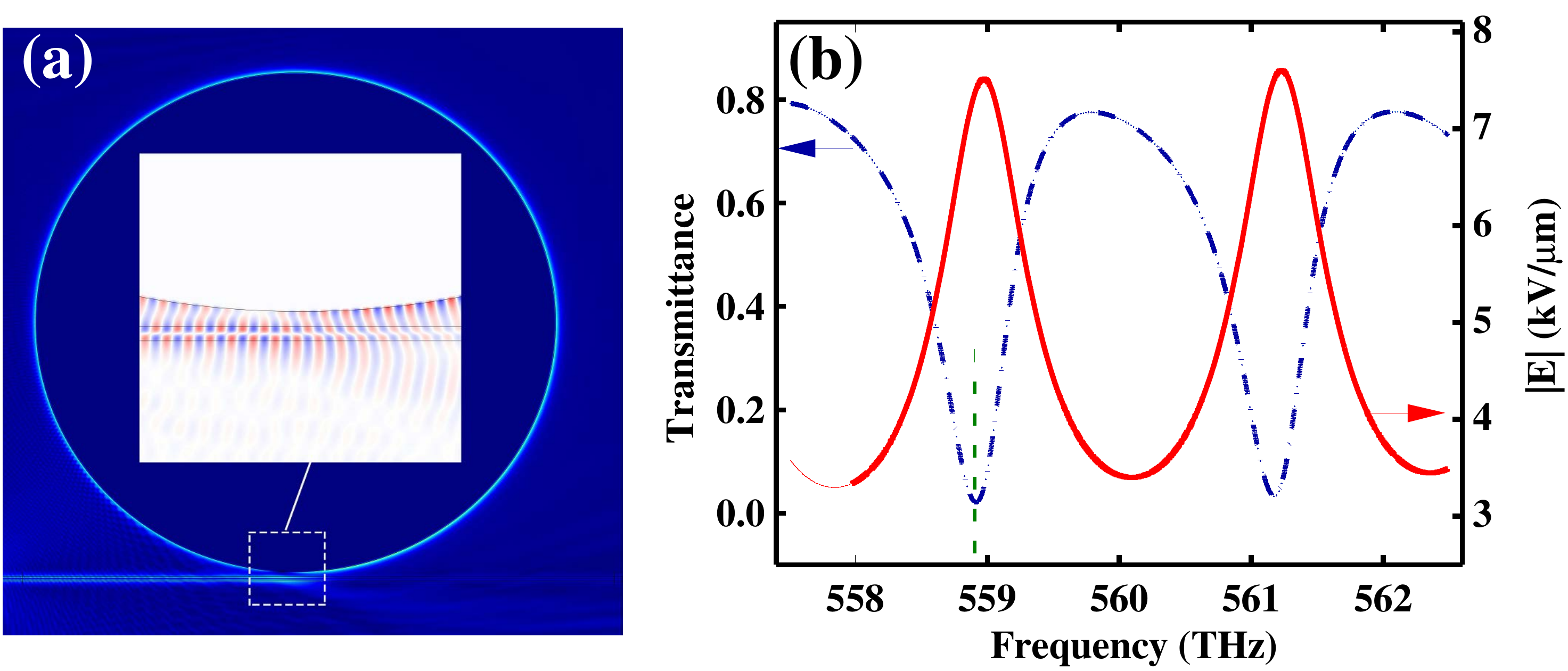}}
\caption{(a) Coupling with second-order guided mode between a straight
silica waveguide and silver cavity, with $R=20\,\mathrm{\protect\mu m}$, and
waveguide width being $0.36\,\mathrm{\protect\mu m}$. (b) Transmission
spectrum of waveguide and the spectrum of normalized energy $\left\vert
E\right\vert $, with $d_{0}=0.36\,\mathrm{\protect\mu m}$.}
\end{figure}

Excitation with the second-order dielectric mode in a straight waveguide is
also investigated, as displayed in Fig. 5(a). TM mode with input power
around $1\,\mathrm{\mathrm{\mu }W/m}$ excites the waveguide at the facet.
Here, we study the influence of second-order dielectric mode only, since the
mismatching between dielectric fundamental mode and plasmon mode is really
large and the inefficient coupling with fundamental mode has been shown in
Fig. 1(c). The inset in Fig. 5(a) is the zoom-in view of the coupling
between dielectric and plasmon modes, where the second-order mode in silica
waveguide is clearly coupled into the silver cavity with minor energy
leaked. Similarly, the transmission spectrum of waveguide and the energy
spectrum of cavity are also calculated, as shown in Fig. 5(b). The
transmission approaching zero at $f=558.9\,\mathrm{THz}$ (dashed line in
Fig. 5(b)) indicates that the coupling between silica waveguide and silver
cavity can even reach critical coupling. And the Q factor of cavity
extracted from the energy spectrum is $Q=1078$ at $f=561.2\,\mathrm{THz}$.

\begin{figure}[tbp]
\centerline{ \includegraphics[width=0.8\columnwidth,natwidth=900,natheight=700]{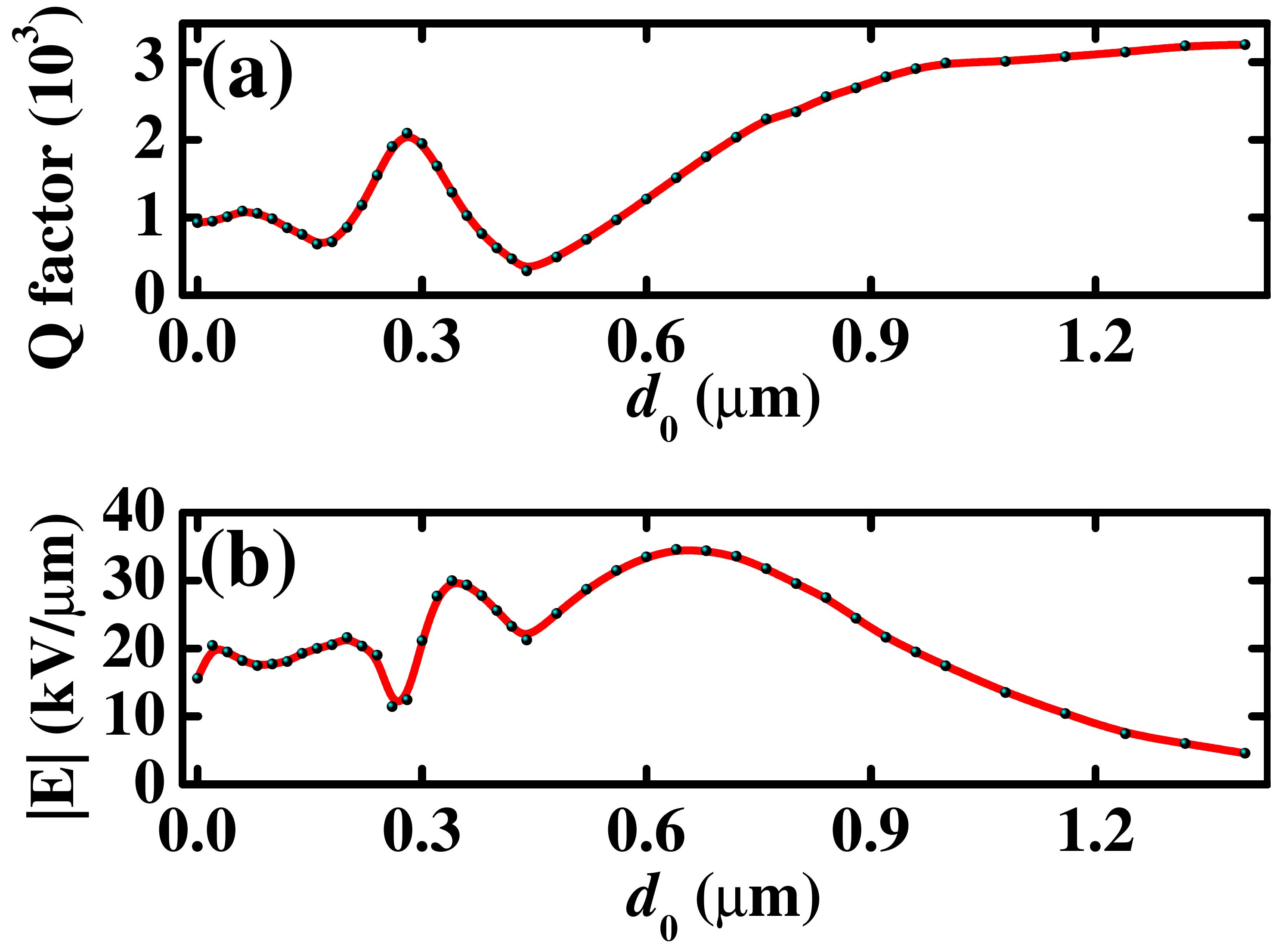}}
\caption{Dependence of Q factor (a) and dependence of normalized energy $%
\left|E\right|$ (b) on $d_{0}$.}
\end{figure}

Then, the performance of this coupling protocol dependent on the minimum gap
between waveguide and cavity $d_{0}$ is studied. The dependences of Q factor
and energy $\left\vert E\right\vert $ on $d_{0}$ are shown in Figs. 6(a) and
6(b) respectively, which are obtained by tracking a specific peak whose
eigen-frequency blueshifts with increasing $d_{0}$. When $d_{0}<0.45\,%
\mathrm{\mathrm{\mu }m}$, the Q factor and $\left\vert E\right\vert $ show
strange oscillation against $d_{0}$. In this regime, the waveguide is very
close to the metal, so the dielectric mode is strongly effected by the
plasmon mode, leading to complex behavior of the integration in Eq. (\ref%
{eq:2}). In addition, $n_{eff}$ of the dielectric mode is close to the
refractive index of air, thus the coupling will also induce loss into the
leaking modes in air. When $d_{0}$ keeps on going up ($d_{0}>0.45\,\mathrm{%
\mathrm{\mu }m}$), the overlap between dielectric and plasmon modes is so
weak that the perturbative coupled-mode theory can be applied. So, $\kappa
_{1}$ exponentially decreases with increasing $d_{0}$, while the cavity
intrinsic loss remains constant. At $d_{0}=0.64\,\mathrm{\mathrm{\mu }m}$, $%
Q\,_{(d_{0}=0.64\,\mathrm{\mathrm{\mu }m})}=\frac{1}{2}Q\,_{(d_{0}=\infty
)} $ (Fig. 6(a)) indicates that the external coupling loss $\kappa _{1}$ is
reduced to equal the intrinsic loss $\kappa _{0}$, corresponding to a
maximum cavity field in Fig. 6(b).

\subsection{Discussion}

From above results, we see that breaking the adiabatic condition effectively
enhances the energy transfer between dielectric waveguide and plasmonic
cavity. For the bent coupler, the efficiency is slightly lower than the case
of higher-order mode coupling, but with the merits of simple excitation with
fundamental mode and compact coupler size. It is well known that the
application of plasmon modes is strongly limited by the Ohmic loss, so
efficient coupling can be expected by reducing the device size to minimize
the loss. Therefore, a bent coupler will be an effective way to excite and
collect plasmon modes. For higher-order mode coupling, it shows critical
coupling when the waveguide is not very close to the cavity. The achieved
cavity field is about 2 times stronger than that for bent coupler. In
addition, the result indicates that higher coupling strength can be achieved
for smaller gap. That is to say, critical coupling is possible for low-Q
exterior WGMs in smaller cavity or cavity with larger metal loss. It's also
shown that the coupling is not very sensitive to the gap, as the field for
contact coupling case ($d_{0}=0$) is about $1/2$ for the optimal case ($%
d_{0}=0.64\,\mathrm{\mathrm{\mu }m}$).

\section{Conclusions}

In summary, we study the coupling between dielectric waveguide modes and
exterior plasmonic whispering gallery modes, which is quite different from
the coupling between dielectric modes. Based on the theoretical analysis of
the principle therein, we propose two protocols to strengthen the
dielectric-plasmon mode coupling. The first one is to break the adiabatic
conversion with a U-shaped bent waveguide, and the other one is to render
the dielectric mode of higher-order to reach phase matching with plasmon
mode. Through numerical simulations, the validity of our proposals is fully
vindicated by the transmission spectra of waveguide and the energy spectra
of cavity. Besides, the influencing factors in two schemes are all analyzed
in detail. Both of the schemes demonstrate that the coupling efficiency can
be greatly raised. The effective coupling between dielectric guided mode and
plasmon mode may find various applications for tunable integrated optical
devices and bio-sensors.

\section*{Acknowledgments}

Chen-Guang Xu and Xiao Xiong contribute equally to this article. This work
was funded by NBRP (grant nos. 2011CBA00200 and 2011CB921200), the
Innovation Funds from the Chinese Academy of Sciences (grant no. 60921091),
NNSF (grant nos. 10904137, 10934006 and 11374289), the Fundamental Research
Funds for the Central Universities (grant no. WK2470000005), and NCET.

\end{document}